\documentclass[lettersize,journal]{IEEEtran}
\usepackage{amsmath,amsfonts,amssymb}
\usepackage{algorithm}
\usepackage{array}
\usepackage[caption=false,font=normalsize,labelfont=sf,textfont=sf]{subfig}
\usepackage{textcomp}
\usepackage{stfloats}
\usepackage{url}
\usepackage{verbatim}
\usepackage{graphicx}
\usepackage{cite}
\usepackage{multirow}%
\usepackage{amsthm}%
\usepackage{mathrsfs}%
\usepackage{xcolor}%
\usepackage{textcomp}%
\usepackage{manyfoot}%
\usepackage{booktabs}%
\usepackage{algorithmicx}%
\usepackage{algpseudocode}%
\usepackage{listings}%
\usepackage{hyperref}%
\usepackage{adjustbox}%
\usepackage{pifont}%

\begin{document}

\title{FOL$\bullet$AI: Synchronized Foley Sound Generation\\with Semantic and Temporal Alignment}

\author{\IEEEauthorblockN{Riccardo~F.~Gramaccioni$^{\flat*}$, Christian~Marinoni$^{\flat*}$, Emilian~Postolache$^{\sharp}$, Marco~Comunità$^{\natural}$, Luca~Cosmo$^{\sharp}$, Joshua~D.~Reiss$^{\natural}$, and~Danilo~Comminiello$^{\flat}$}

\IEEEauthorblockN{\textit{$^{\flat}$Sapienza University of Rome, Italy}\\\textit{$^{\sharp}$Ca' Foscari University of Venice, Italy}\\\textit{$^{\natural}$Queen Mary University of London, UK}
}}


\maketitle

\begin{abstract}
Traditional sound design workflows rely on manual alignment of audio events to visual cues, as in Foley sound design, where everyday actions like footsteps or object interactions are recreated to match the on-screen motion. This process is time-consuming, difficult to scale, and lacks automation tools that preserve creative intent. 
Despite recent advances in vision-to-audio generation, producing temporally coherent and semantically controllable sound effects from video remains a major challenge. To address these limitations, we introduce Fol·AI, a two-stage generative framework that decouples the \textit{when} and the \textit{what} of sound synthesis, i.e., the temporal structure extraction and the semantically guided generation, respectively. In the first stage, we estimate a smooth control signal from the video that captures the motion intensity and rhythmic structure over time, serving as a temporal scaffold for the audio. In the second stage, a diffusion-based generative model produces sound effects conditioned both on this temporal envelope and on high-level semantic embeddings, provided by the user, that define the desired auditory content (e.g., material or action type). This modular design enables precise control over both timing and timbre, streamlining repetitive tasks while preserving creative flexibility in professional Foley workflows. Results on diverse visual contexts, such as footstep generation and action-specific sonorization, demonstrate that our model reliably produces audio that is temporally aligned with visual motion, semantically consistent with user intent, and perceptually realistic. These findings highlight the potential of Fol·AI as a controllable and modular solution for scalable, high-quality Foley sound synthesis in professional and interactive settings. Supplementary materials, including samples and code, are accessible on our dedicated demo page at \url{https://ispamm.github.io/FolAI}.
\end{abstract}

\newcommand{\cem}[1]{\textcolor{blue}{cem: #1}}
\section{Introduction}
\label{sec:intro}

\begin{figure}[t!]
    \centering
    \includegraphics[width=\linewidth]{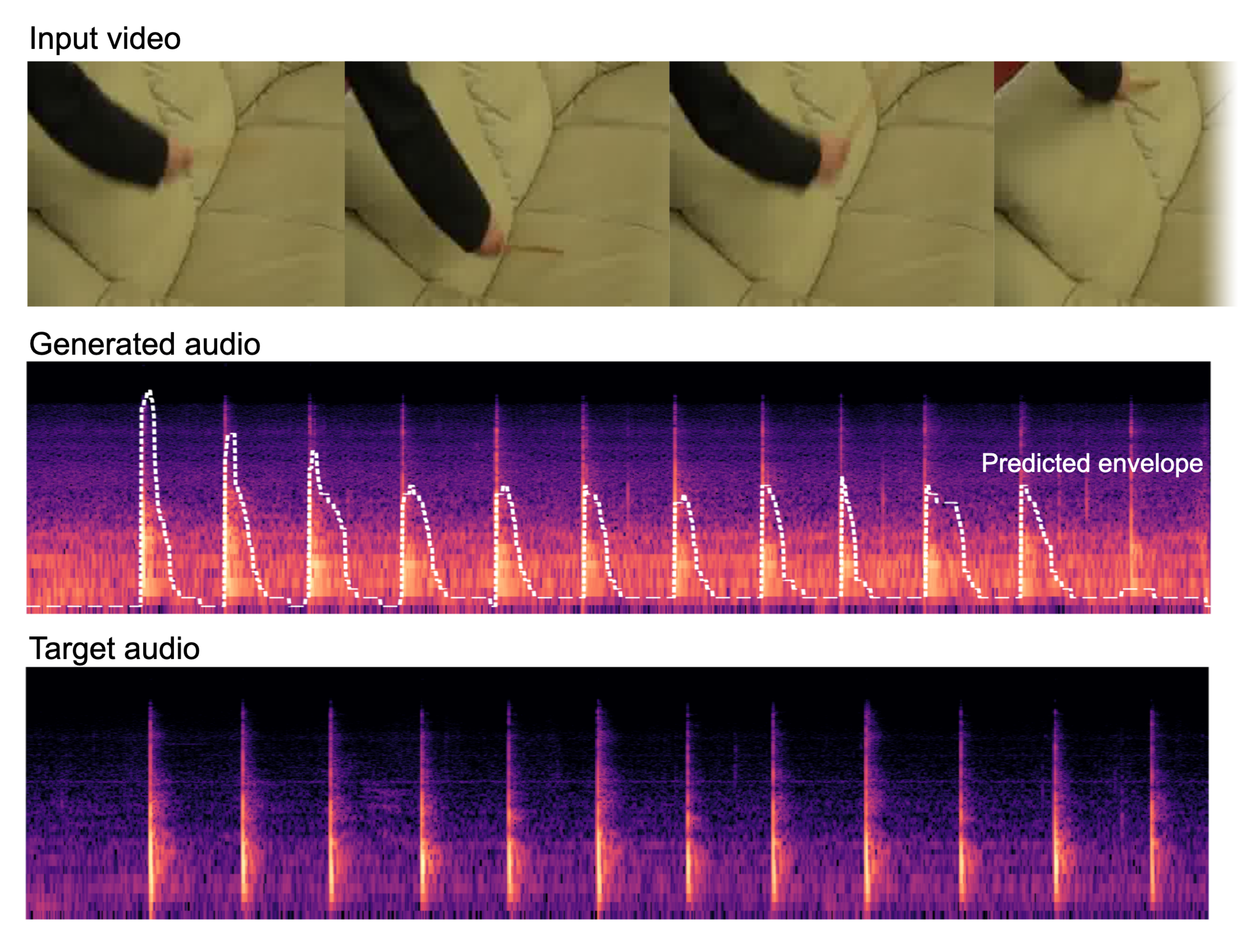}
    \caption{Example showing ground truth audio and video, predicted RMS envelope and generated audio.}
    \label{fig:risultati-spettrogrammi}
\end{figure}

Sound is a fundamental component of audiovisual storytelling, capable of deeply altering the meaning and emotional impact of a scene. Traditionally, Foley artists and sound designers manually select, record, and synchronize sounds (like footsteps or object interactions) to match on-screen actions. This careful process ensures the desired narrative effect but is inherently time-consuming and requires significant effort. Recent advancements in deep learning, particularly in multimodal generative models \cite{Suzuki2022ASO, Guizzo2021L3DAS21CM, Guizzo2022L3DAS22CL} and Large Language Models (LLMs) \cite{Zhang2023VideoLLaMAAI, tang2024empoweringllmspseudountrimmedvideos}, have generated significant interest in automating and enhancing audio synthesis. This has led to the rise of Video-to-Audio (V2A) generation \cite{aytar2016soundnet, Zhou2017VisualTS, Sheffer2022IHY, 9126216, 9782577}, a research area focused on synthesizing audio that is semantically relevant and temporally aligned with an input video sequence. Initial efforts in this area often focused on semantic alignment, generating ambient sounds or soundtracks reflecting the overall mood \cite{di2021video, zhuo2023video, Chen2024SemanticallyCV, Gao2023AnOO}, but lacked the temporal precision needed for synchronizing specific sound effects.

Achieving both appropriate semantic content and precise timing is crucial for effective V2A generation as temporal alignment is equally fundamental for realism, particularly when synchronizing distinct sound events like Foley effects. Even slight timing mismatches, on the order of milliseconds, can be perceived by the audience and disrupt the feeling of immersion \cite{keetels2012perception}. Furthermore, although more recent end-to-end V2A models \cite{Zhou2017VisualTS, 9782577, Chen2018VisuallyIS, Cui2022VarietySoundTV, Yi2024EfficientVT, Ishii2024ASB} aim for better synchrony, they often function as ``black boxes". This typically offers limited direct control for sound designers to refine the timing or creatively adjust the generated audio beyond basic post-processing. This lack of fine-grained control hinders their adoption in professional workflows where artistic input and adjustments are essential. 

\begin{figure*}[!t]
    \centering
    \includegraphics[width=\textwidth]{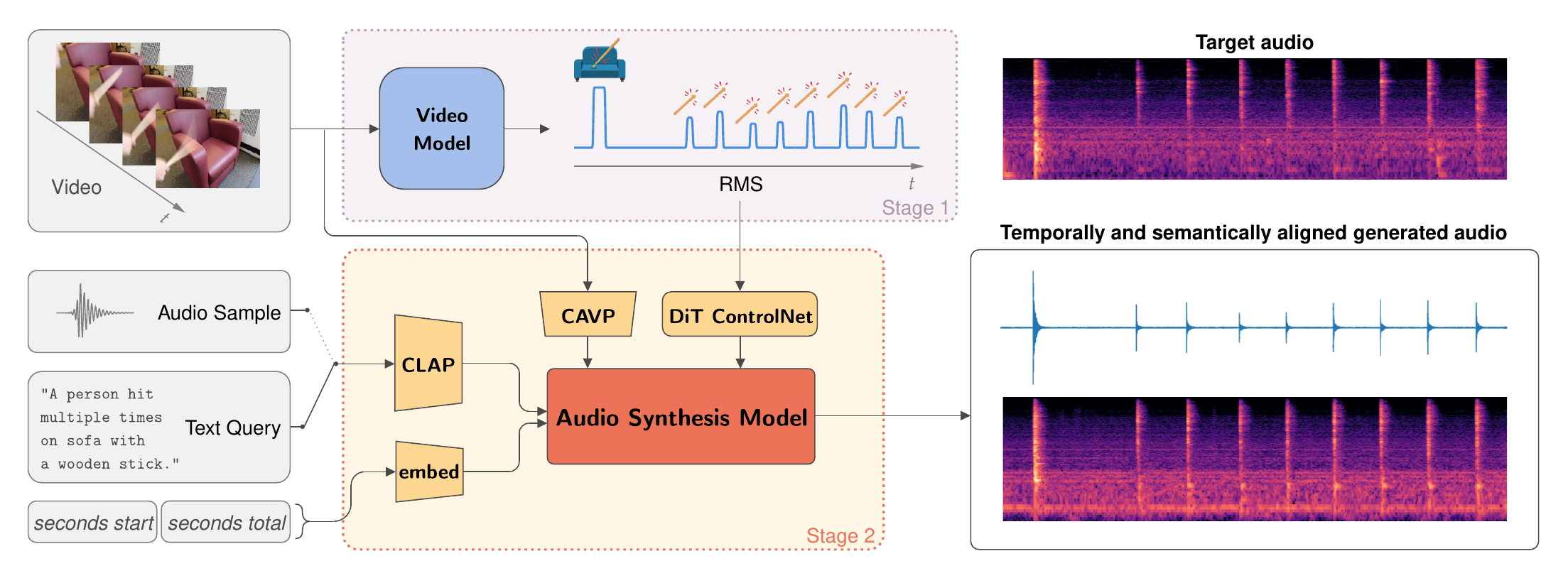}
    \caption{FOL·AI architecture consists of two distinct parts: the video model, that predicts an envelope representative for the audio directly from the input video, and the audio synthesis model for the controlled generation of the final audio effect. The generation is controlled temporally by the predicted RMS envelope through a DiT ControlNet, and semantically by CLAP and CAVP embeddings. The length of the output waveform can be controlled with $\mathrm{seconds\_start}$ and  $\mathrm{seconds\_total}$ parameters.}
    \label{fig:stablev2a-architecture}
\end{figure*}

To address these critical needs for robust temporal alignment and user control, we introduce FOL·AI, a novel two-stage generative framework designed for accurate V2A synthesis. Our approach separates the task into predicting the temporal structure and then generating the audio based on that structure and desired sound characteristics, as shown in Fig. \ref{fig:stablev2a-architecture}.
The first stage analyzes the input video using frame features and optical flow to predict a Root Mean Square (RMS) envelope. This envelope represents the audio's timing, intensity, and duration \cite{Chung2024TFoleyAC}. Importantly, this predicted envelope is human-readable and can be directly edited by the sound designer. This allows for fine-tuning the timing or even adding sounds for off-screen events before the audio is generated.
The second stage employs a state-of-the-art diffusion model architecture to create the final sound. To ensure the audio matches the timing specified by the envelope, we introduce an additional conditioning network architecture that steers the diffusion process using the RMS envelope from the first stage. Users can control the timbre by providing semantic information through embeddings from reference audio samples, text descriptions and the input video. This modular design allows creatives to control \textit{when} sounds occur and \textit{what} they sound like.

The FOL·AI framework successfully generates 44.1 kHz stereo audio of variable length that shows strong temporal and semantic synchronization with the input video, as qualitatively illustrated in Fig. \ref{fig:risultati-spettrogrammi}. We demonstrate the effectiveness of our model through careful objective evaluations on the widely used \textit{Greatest Hits} V2A benchmark dataset \cite{Owens2015VisuallyIS}, where FOL·AI achieves state-of-the-art results compared to existing V2A models.
Furthermore, recognizing the lack of suitable public datasets with high-quality audio and video for evaluating V2A models on specific, highly relevant sound design tasks like Foley, we introduce \textit{Walking The Maps}. This new dataset focuses specifically on footstep generation, a key challenge and frequent task in the field \cite{comunita2022neural}. Sourced from high-quality video game walkthroughs available online, it provides numerous video clips featuring clearly audible footstep sounds across various surfaces (like grass, concrete, wood) and character actions (walking, running), paired with corresponding high-definition video offering excellent temporal cues. The audio was carefully preprocessed using source separation techniques to isolate footstep sounds, creating clean targets suitable for robust V2A model training and evaluation. We make \textit{Walking The Maps} freely available to the research community to facilitate progress and benchmarking in realistic automated Foley synthesis scenarios.

\section{Related Works}
\label{sec:format}

Generating audio that is aligned with a silent reference video is salient task nowadays for multimedia post production. Due to recent developments in deep learning, many of the leading experts in the audiovisual industry are focusing their efforts on adopting deep learning models that can be integrated into the principal post production tools \cite{Ceylan2023Pix2VideoVE, Jenni2023AudioVisualCL, Fedorishin2024AudioMC, sahipjohn24_interspeech}.

Despite these advancements, generating audio from video still presents many challenges. 
Not only the audio produced must maintain semantic coherence with what is shown in the video, but generated sounds also have to exhibit enough time alignment with the actions throughout the video to allow an adequate sense of realism while watching the scene. Although several audio models manage to generate reasonably realistic sounds for Foley synthesis \cite{Chung2024TFoleyAC, Colombo2024MambaFoleyFS, Liu_MLP2022_01, yuan2023latent, Lee2020RealTimeSS}, many V2A models struggle when it comes to actually achieving good time alignment.

Im2Wav \cite{Sheffer2022IHY} focuses on generating sounds that are semantically relevant to an image or sequence of images, using CLIP \cite{Radford2021LearningTV} to condition with visual features two transformer language models.
In order to perform a cross-modality generation between video and audio, Xing, et al. \cite{Xing2024SeeingAH} propose a multimodality latent aligner with the pre-trained ImageBind \cite{Girdhar2023ImageBindOE} model, used to condition a latent diffusion model (LDM) in order to generate an audio that is semantically relevant for an input video. 

RegNet \cite{Chen2020GeneratingVA} represents one of the first attempts to achieve semantic as well as temporal synchronization. 
This model uses a simple and efficient video encoder that extracts relevant visual features from frames and optical flow of an input video and uses them to condition a Generative Adversarial Network (GAN) to generate visually aligned sounds. 
SpecVQGAN \cite{SpecVQGAN_Iashin_2021} also uses RGB and optical flow features of a video but leveraging a more powerful Transformer-based autoregressive architecture to generate temporally and semantically synchronized sounds to an input video.

A similar architecture is used in CondFoleyGen \cite{du2023conditional}, where such a model is trained directly using \textit{Greatest Hits}, succeeding in achieving an efficient alignment in both content and timing with the reference video.
Instead, Diff-Foley \cite{NEURIPS2023_98c50f47} uses Contrastive Audio-Visual Pretraining (CAVP) to temporally and semantically align audio and video modalities, being able to generate a video embedding that contains features relevant to the corresponding audio. Such features are used to directly condition an LDM. 
However, all these models do not provide human-intelligible control, thus not allowing direct supervision by sound designers over the final generation, as they cannot act on either the semantics or the timing of the final output.

SyncFusion \cite{Comunit2023SyncfusionMO} - which was the first to introduce a human readable control for the V2A task - uses a video encoder based on a ResNet(2+1)D-18 \cite{Tran2017ACL} that, by taking frames of the target video as input, generates an onset track; this onset track is then passed to a time-domain diffusion model \cite{schneider2023mo} to generate the final output. A similar approach is implemented in \cite{10890132}. The visual representation provided by the onsets is very similar to the manual annotations used by sound designers when determining the temporal placements of the sound sources to be sonorized in a video. Therefore, it provides an informative and yet easy-to-edit control for the user, allowing direct oversight on the generation of the final output. This onset track is represented by a binary mask that indicates the presence or absence of the action of interest for each frame of the video. Hence, an onset can indicate the temporal location of a sound but can not inform either about the intensity or temporal duration.
Furthermore, the onset of a sound event cannot be determined unambiguously, as in the case of a sound event sustained over time where the temporal placement of the onset is not unequivocally determined. Consequently, in order to use onsets, it is necessary to have a dataset containing manual annotation for every sound event in each video.

T-Foley \cite{Chung2024TFoleyAC}, although not working with an input video, uses an envelope to condition a diffusion model similar to the one in SyncFusion, demonstrating the effectiveness of such a control for generating highly temporally conditioned audio. This model can generate sounds that follow with high accuracy the temporal guidance provided by the envelope, extending this idea in Video-Foley \cite{Lee2024VideoFoleyTV}, where AudioLDM \cite{Liu2023AudioLDMTG} is employed to generate 16 kHz mono audio aligned temporally and semantically with the input video.

\section{Background}
\subsection{Audio Diffusion Models}
Diffusion Models currently represent the state-of-the art in generative deep learning. They are based on non-equilibrium thermodynamics and are defined by a Markov chain of diffusion steps that slowly add noise to target data. The aim of such models is to reverse this process in order to learn how to generate data of interest from noise. 

Denoising Diffusion Probabilistic Models were introduced in \cite{Ho2020DenoisingDP}. 
They were first used to obtain state-of-the-art results in image generation and then in the synthesis of different media, such as video and audio.
Like Diffusion Models for images, Audio Diffusion Models (ADM) start by producing a random, noisy audio signal and gradually enhance it through multiple refining iterative steps. In each step, the noise is reduced, and finer details are added to the audio. 
Such models have been used to generate both music \cite{mariani2024multisource} and environmental sounds \cite{kong2021diffwave}.

Latent Diffusion Models \cite{Rombach2021HighResolutionIS} perform the diffusion process in the latent space. In this case, high-dimensional data of interest $\mathbf{y}$, such as audio, are encoded in low-dimensional latent embeddings $\mathbf{z} = \mathcal{E}(\mathbf{y})$, providing a more meaningful representation for training the network. For V2A, the process is guided by a set of conditionings $\mathbf{C} = \mathbf{c}_1, \mathbf{c}_2, ..., \mathbf{c}_n$.
In the forward process, Gaussian noise is slowly added to the original data distribution with a fixed schedule $\alpha_1, \ldots, \alpha_T$, where $T$ is the total timesteps, and $\bar{\alpha}_t = \prod_{i=1}^{t} \alpha_i$:

\begin{equation}
q(\mathbf{z}_t | \mathbf{z}_{t-1}) = \mathcal{N}(\mathbf{z}_t; \sqrt{\alpha_t} \mathbf{z}_{t-1}, (1 - \alpha_t) \mathbf{I})
\end{equation}
\begin{equation}
q(\mathbf{z}_t | \mathbf{z}_0) = \mathcal{N}(\mathbf{z}_t; \sqrt{\bar{\alpha}_t} \mathbf{z}_0, (1 - \bar{\alpha}_t) \mathbf{I}).
\end{equation}

The model should attempt to reverse the process by optimizing a denoising objective, typically defined as:

\begin{equation}
\mathcal{L}_{\text{LDM}} = \mathbb{E}_{\mathbf{z}_0, t, \epsilon} \| \epsilon - \epsilon_\theta(\mathbf{z}_t, t, \mathbf{C}) \|_2^2,
\end{equation}

where $\epsilon$ is Gaussian noise and $\epsilon_\theta(\mathbf{z}_t, t, \mathbf{C})$ denotes the estimated noise, which is the output of the model.

After training, LDMs generate latents by sampling through the reverse process with $\mathbf{z}_T \sim \mathcal{N}(0, \mathbf{I})$ formulated as:

\begin{equation}
p_\theta(\mathbf{z}_{t-1} | \mathbf{z}_t) = \mathcal{N}(\mathbf{z}_{t-1}; \mu_\theta(\mathbf{z}_t, t, \mathbf{C}), \sigma_t^2 \mathbf{I})
\end{equation}

\begin{equation}
\mu_\theta(\mathbf{z}_t, t, \mathbf{C}) = \frac{1}{\sqrt{\alpha_t}} \left( \mathbf{z}_t - \frac{1 - \alpha_t}{\sqrt{1 - \bar{\alpha}_t}} \epsilon_\theta(\mathbf{z}_t, t,  \mathbf{C}) \right)
\end{equation}

\begin{equation}
\sigma_t^2 = \frac{1 - \bar{\alpha}_{t-1}}{1 - \bar{\alpha}_t} (1 - \alpha_t).
\end{equation}

Finally, the desired output  $\hat{\mathbf{y}}$ is obtained by decoding the generated latent $\mathbf{z}_0$ with a decoder $\mathcal{D}$.

These models are widely used for audio generation \cite{Liu2023AudioLDM2L}, with Stable Audio \cite{Evans2024FastTL, Evans2024LongformMG} representing a state-of-the-art latent ADM.

\subsection{Guiding diffusion processes with ControlNet}

ControlNet \cite{zhang2023adding} is a neural network specifically designed for diffusion models that allows more precise control over the generation process by conditioning the model on additional inputs. It was first used to control image generation in Stable Diffusion with extra information such as features or prompts.
Initially designed for U-Net-based diffusion models, ControlNet was later extended in \cite{chen2024pixartalpha} to work with Diffusion Transformers (DiT) for image synthesis. 

In audio diffusion models, ControlNet can guide the generation of sound by incorporating specific constraints like tempo, rhythm, pitch, or other audio characteristics \cite{Wu2023MusicCM}. This makes it useful for tasks like generating audio that adheres to a desired structure or style, improving the accuracy and flexibility of the diffusion process.

\section{Proposed Method}

\subsection{Problem Formulation}

Let us consider a target video $\mathbf{x} \in \mathbb{R}^{T \times C \times H \times W}$, where $T$ is the total duration of the video expressed in frames, $C$ is the number of input RGB channels, i.e. 3 channels, $H \times W$ is the dimension in height and width of each frame. The objective of the proposed model is to generate an audio track $\mathbf{y} \in \mathbb{R}^{Ch \times L}$, where $Ch$ represents the number of audio channels, that is 2 for stereo audio, and L is the time duration of the audio expressed in samples. The generated audio $\mathbf{y}$ must be semantically and temporally synchronized with the input video, so it can be used as a realistic soundtrack for it. 
For example, in the case of a video depicting a person walking slowly on a wooden pavement, the audio generated by the model must be both in terms of semantics and temporal alignment distant from the audio generated for a video depicting a person running on grass. 

One of the main challenges when approaching V2A problems is the completely different temporal resolution between video and audio. Indeed, the temporal resolution of a video, expressed in frames per second (\textit{fps}), is usually much lower than the temporal resolution of an audio, expressed in sample rate (\textit{sr}).
For instance, a commonly used frame rate in video is 30 \textit{fps}, while the standard in audio is represented by a \textit{sr} of 44.1 kHz. This means that in this specific case 1470 audio samples are represented by one single frame.
One possible solution to mitigate this difference is to increase the \textit{fps} of the video and lower the \textit{sr} of the audio. 
However, increasing the frame rate of a video excessively leads to a significant rise in the computational cost of the model. On the other hand, decreasing the sample rate results in a loss of audio quality, which is unsuitable for practical use in sound design.
It is therefore crucial to carefully manage the transition between the two modalities, i.e., audio and video, to ensure that the problem remains computationally feasible while maintaining high enough audio quality to capture all the features of the sound needed by sound designers and Foley artists \cite{garcia2024discerning, zong2024a}.

\subsection{Mapping a Video to an Envelope}

In audio, the envelope describes how a sound evolves over time. It is represented by a smooth curve outlining the extremes of the signal, providing useful information such as its amplitude and duration over time. Moreover, by tracking the shape of the waveform, it is an effective and visually understandable feature for indicating the temporal locations of all sound events in the audio. 

The envelope can be calculated in different manners. In T-Foley \cite{Chung2024TFoleyAC}, the Root-Mean-Square (RMS) of the waveform, a commonly used frame-level amplitude envelope feature, is employed for temporal-event guided waveform generation.
The $i$-th sample of the temporal sequence representing the RMS envelope is then calculated on a window of the audio signal $\mathbf{y}$ as follows:

\begin{equation}
    \mathbf{r}_i = \mathbf{RMS}_i(\mathbf{y)} = \sqrt{\frac{1}{W} \sum_{t=ih}^{ih+W}\mathbf{y}^{2}(t)},
\label{eqn:RMS}
\end{equation}
where $W$ is the window size and $h$ is the hop size.

We address this task as a classification one since trying to map a curve representing the envelope of an audio as a regression task does not produce satisfactory results \cite{Lee2024VideoFoleyTV}. 
The reason is that sounds related to actions are often transient, which makes audio containing multiple repeated actions sparse, i.e. most samples represent silence. A model that tries to map such a time series as a regression task thus tends to predict silence or the mean value of the curve, as we also found out in our early experiments. 

We then perform amplitude quantization of the envelope by $\mu$-law encoding. This encoding is widely used in audio, especially in pitch estimation \cite{Oord2016WaveNetAG}, precisely to mitigate the transient and sparse nature of waveform of this kind. This encoding allows amplitude quantization of the signal by dividing the values into equidistant bins. 
The idea here is to map video frames to a sequence of specific classes, representing the amplitude values of the envelope.


\begin{figure}[t]
    \centering
    \includegraphics[width=0.87\linewidth]{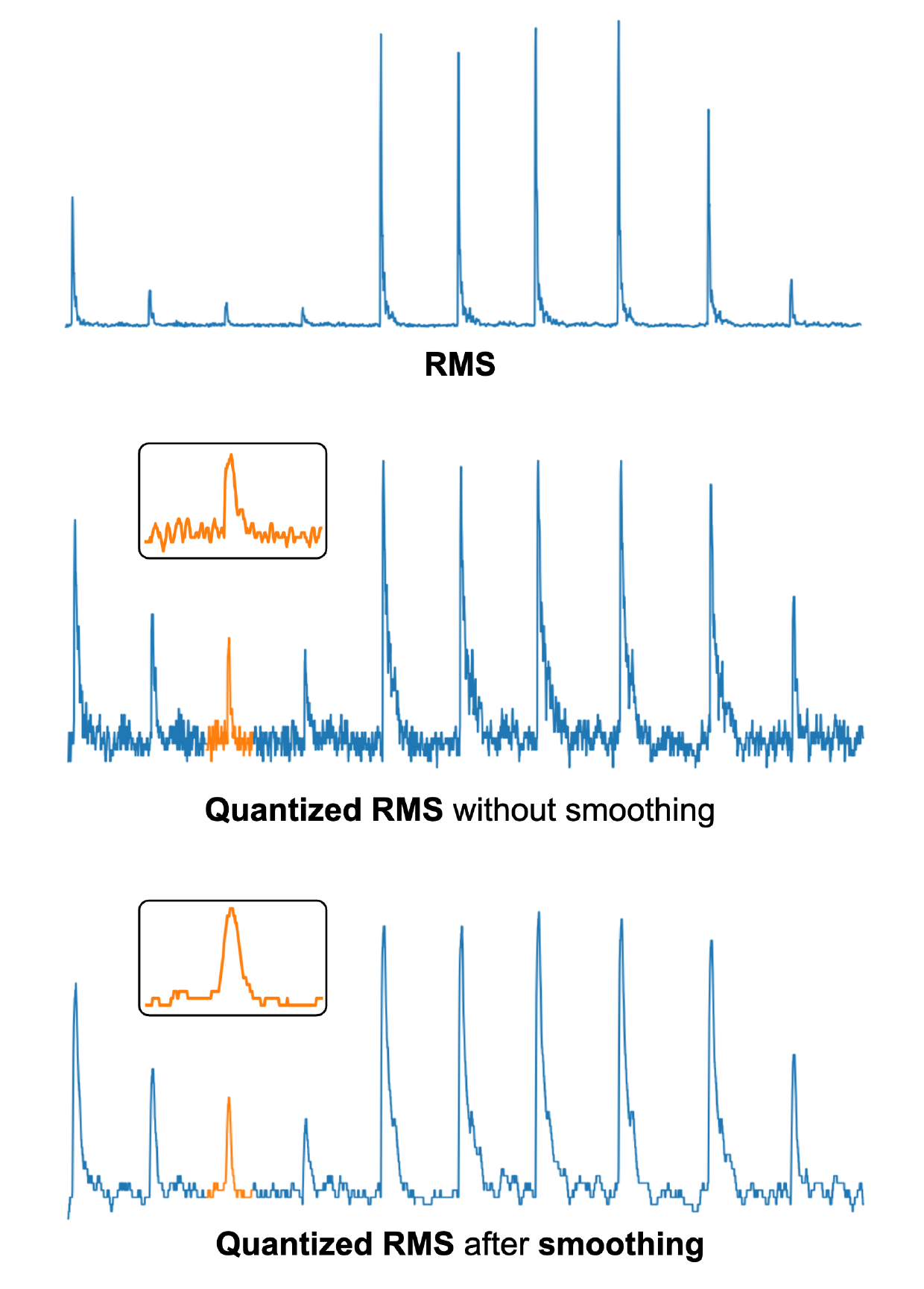}

    \caption{Example of the ground truth envelope before and after the smoothing operation.}
    \label{fig:rms}
\end{figure}


Furthermore, it is necessary not to penalize near-correct predictions. Indeed, a precise prediction of the correct class is desirable but not strictly required to ensure good perceptual quality.
Then, in order to minimize the penalty of a near-correct class prediction, Gaussian label smoothing on the target envelopes - as proposed in \cite{Kum2019JointDA} - is used:
\begin{equation}
\mathbf{r}(i) = 
\begin{cases}
\exp\left(-\frac{(c_i - c_{\text{gt}})^2}{2\sigma^2}\right) & \text{if } |c_i - c_{\text{gt}}| \leq W \, (c_i, c_{\text{gt}} \neq 0) \\
0 & \text{otherwise},
\end{cases}
\end{equation}
where $i$ is the class index, $c_{\text{gt}}$ is the ground-truth class, $\sigma = 1$, and $W$ is the window size on whose values the smoothing is applied.

\subsection{Video Model}

The first stage of our model, the video model, is a simple neural network that takes some features extracted from the video as input and generates a temporal guide for the sound synthesis process of the second stage.

The target is an envelope extracted from a corresponding 10-second audio chunk sampled at 16 kHz. This relatively low sample rate helps reduce the temporal resolution gap between audio and video while preserving sufficient detail in the audio track.
We compute the RMS with the function implemented in the Librosa Feature module\footnote{\url{https://librosa.org/doc/main/generated/librosa.feature.rms.html}}. Following the nomenclature of Eq. \ref{eqn:RMS}, we set $W=512$ and $h=128$, thus obtaining an envelope $\mathbf{r}$ with 1250 time samples. Since we normalize the audio $\mathbf{y}$ in the $[-1, 1]$ range, the resulting $\mathbf{r}$ is a curve with values in the range $[0, 1]$.

In addition, to mitigate the high-frequency variations which are common in real sound waveforms, we apply a smoothing filter with a kernel size of 15. 
Our experiments show that incorporating such a filter is essential to improve the results, simplifying the classification task that the network performs on the video frames, as can be seen in Fig. \ref{fig:rms}.
Finally, we map the time-continuous RMS-envelope to 64 distinct classes by applying $\mu$-law encoding, using the function provided in the Librosa library\footnote{\url{https://librosa.org/doc/main/generated/librosa.mu_compress.html}}.
We then apply Gaussian label smoothing on the target discretized envelope.

the video model takes as input features extracted from the videos. We first convert each video to 30 \textit{fps}.
This means that a 10-second long video is represented by a sequence of 300 consecutive frames.

In addition, we compute the optical flow of each video using the RAFT \cite{Teed2020RAFTRA} model. 
To effectively capture both appearance and motion cues relevant to sound generation, we utilize features extracted by Temporally Contextualized CLIP (TC-CLIP) \cite{kim2024tcclip} from both RGB frames and RAFT-computed optical flow. TC-CLIP excels at video understanding by integrating temporal context. Our experiments confirm that combining static frame appearance (RGB) with motion information (optical flow) via TC-CLIP significantly improves the ability of the video model to predict accurate audio envelopes, as motion is often a key indicator of sound events.

TC-CLIP generates 512 dimensional frame-wise features.
We concatenate the RGB and optical flow features along the feature dimension, finally obtaining an input of size $[\mathrm{video\_frames}, 1024]$, where in our case $\mathrm{video\_frames} = 300$ for \textit{Greatest Hits} experiments.

Then, we use an efficient model based on the RegNet video encoder \cite{Chen2020GeneratingVA}, 
and we improve it to generate the output of our model. 
The network is composed of three 1D convolutional layers, a two-layer bidirectional LSTM (Bi-LSTM) \cite{Graves2005FramewisePC} and a linear layer that projects the resulting features to the output size required for the classification task. 
We use the convolutional block as an upsample branch needed to scale features from video temporal resolution to the audio temporal resolution. 
Each convolutional layer is followed by a batch normalization layer, a ReLU activation function and an Upsample layer. The upsample sizes for the three layers are $[600, 1200, 1250]$, thus increasing the time dimension of the output from $\mathrm{video\_frames}$ to $\mathrm{RMS\_frames}$, where $\mathrm{RMS\_frames}$ is equal to $1250$. Thanks to this upsample, we do not perform a frame-by-frame classification of the video, but we directly map the continuous RMS envelope. 

\begin{figure}[t]
    \centering
    \includegraphics[width=0.85\linewidth]{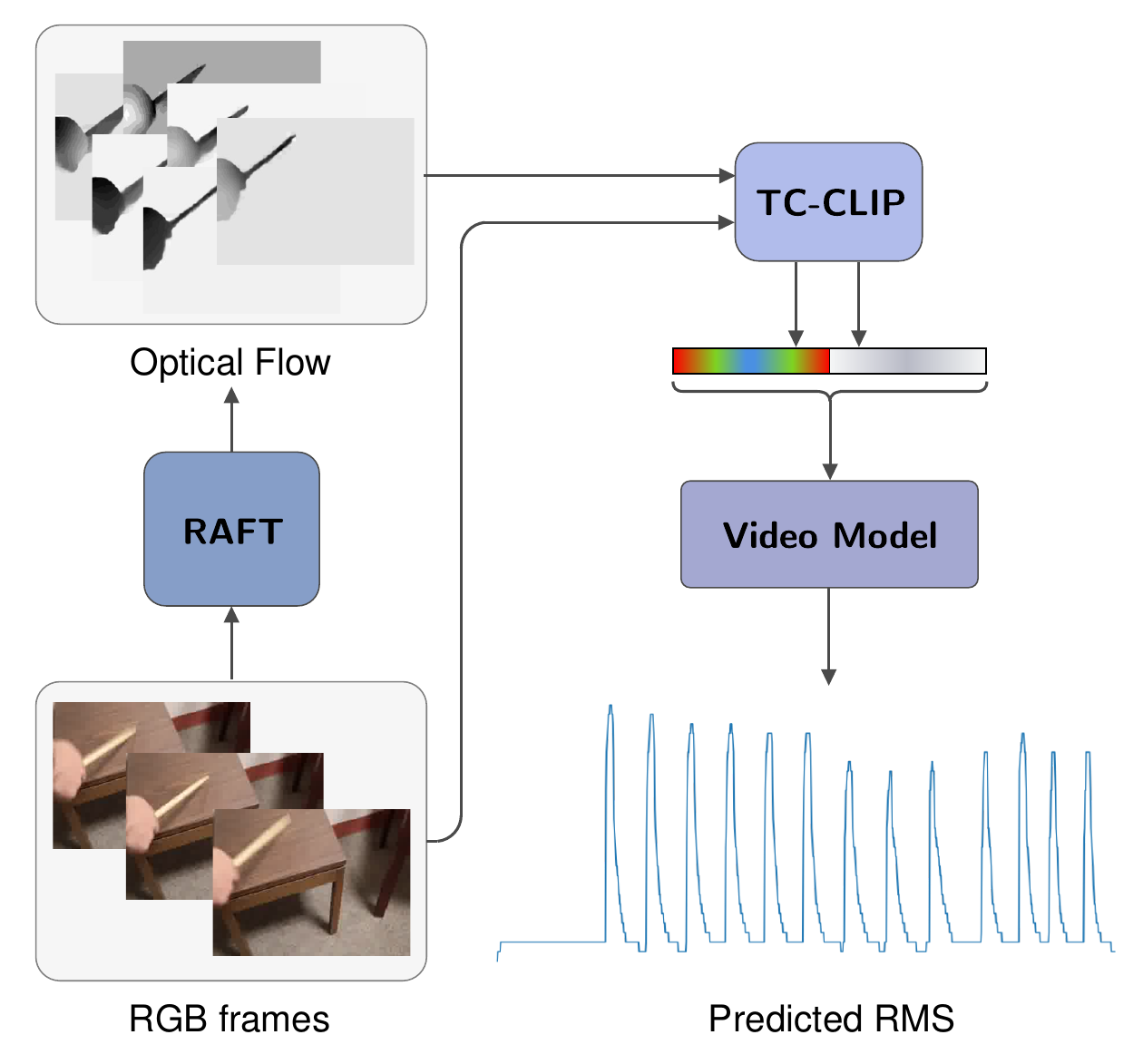}

    \caption{Block diagram for the proposed the video model.}
    \label{fig:rms-mapper}
\end{figure}

We use a Cross Entropy loss as objective for our classification problem:
\begin{equation}
    \mathcal{L}_{\text{CE}}(\mathbf{r}_{\mathbf{d}}, \hat{\mathbf{r}}_{\mathbf{d}}) = -\sum_{i=1}^C {\mathbf{r}_{\mathbf{d}}}_i \log(\hat{\mathbf{r}}_{\mathbf{d}_i}),
\end{equation}
where $\mathbf{r}_{\mathbf{d}}$ is the discretized RMS envelope $\mathbf{r}$, i.e. the envelope after $\mu$-law encoding and Gaussian label smoothing, and $\hat{\mathbf{r}}_{\mathbf{d}}$ is the output of our model. Fig. \ref{fig:rms-mapper} depicts a block diagram of the proposed the video model.

\subsection{Audio Synthesis Model}
The audio synthesis model leverages Stable Audio Open \cite{Evans2024StableAO}, which is an Audio Latent Diffusion Model for generating long-form, variable-length stereo
music and sounds at 44.1 kHz using text prompts. 
Therefore, this model is not specifically trained for V2A tasks, which means that the generated audio does not take into account a time alignment with a specific reference. 

We want to use the prior knowledge of such a state-of-the-art model for audio synthesis, but at the same time guide audio generation both semantically and temporally. This means that the audio synthesis model should be controlled by the envelope extracted from the the video model, representing our temporal control, and conditionings that capture the semantics and duration of the audio. 

The aim of the synthesis model is to learn a probability distribution $p(\mathbf{y} | \mathbf{F}, \mathbf{r})$ of a waveform $\mathbf{y}$, given a time-independent set of conditionings $\mathbf{F}={\mathbf{f}_1, \mathbf{f}_2,..., \mathbf{f}_n}$, representing a set of $n$ desired semantic characterizations,  and $\mathbf{r}$, that is the temporal control.

The audio synthesis model is trained on the same L2 loss on which Stable Audio models are trained.

\subsubsection{Temporal Control}

We use a ControlNet to drive generation using a temporal control and at the same time just fine-tune Stable Audio Open on our specific task, without having to train the model from scratch. 
A key innovation of the audio synthesis model lies in its temporal control mechanism. We adapt the ControlNet framework, originally designed for U-Nets, to operate with  Diffusion Transformer (DiT) architecture of Stable Audio, following recent advancements \cite{chen2024pixartalpha}. To our knowledge, this represents the first successful integration of ControlNet with Stable Audio's DiT to impose explicit, fine-grained temporal control via an external signal like the RMS envelope, allowing us to precisely guide the synthesis process while leveraging the power of the pre-trained Stable Audio model.

We want to generate stereo audio at 44.1 kHz. In the training phase, the length of the audio to be generated is fixed at 10s. Accordingly we interpolate the ground truth RMS envelope, extracted for both L/R channels, to the length in samples of the target audio signal. 

The input of the ControlNet needs to be of the same dimension as the input of the diffusion process, i.e. the embedding of the audio used in Stable Audio. 
To extract this latent representation of the envelope, we use the VAE introduced in \cite{Evans2024FastTL} for encoding the input to the diffusion model, without needing to update its weights. 
This VAE downsamples the input stereo audio by a factor of 1024. Specifically, it maps an input signal $\mathbf{y} \in \mathbb{R}^{2 \times L}$, where $L$ is the number of samples representing the waveform, and $2$ is the number of channels, to an embedding $\mathbf{y_c} \in \mathbb{R}^{64 \times \frac{L}{1024}}$. 

In all of our experiments, such a VAE has proven to provide a meaningful representation of the RMS envelope as well, so we use it to encode the input of the ControlNet  $\mathbf{r}$ and the diffusion process input  $\mathbf{y}$.

\begin{figure}[!t]
    \centering
    \includegraphics[width=\linewidth]{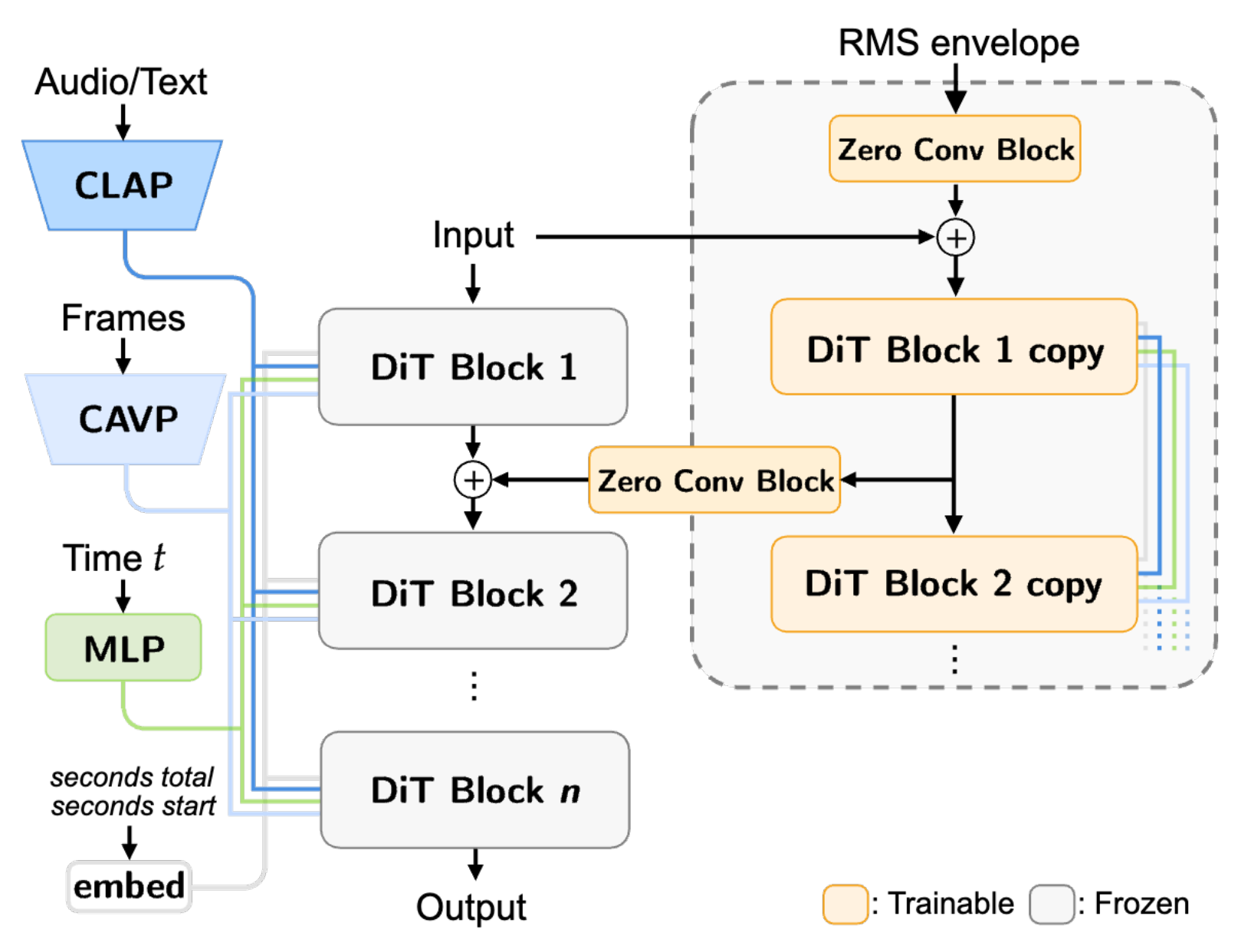}

    \caption{Block diagram for the audio synthesis model. Trainable modules are ControlNet blocks while Stable DiT blocks are frozen.}
    \label{fig:stable-foley}
\end{figure}

For the design of the audio synthesis model we follow the general architecture of ControlNet: the DiT layers are then frozen, and a trainable copy for each of them is created with two zero-initialized convolutional layers placed before and after the copy. For training, the envelope is computed so that it has the same length in samples as the target audio. The control signal $\mathbf{r}$, representing the RMS envelope, and the input waveform $\mathbf{y}$ are both encoded through the same VAE. For each layer, the encoded control signal $\mathbf{r_c}$ is first processed through the first zero-initialized convolutional layer, then added to the input $\mathbf{y_c}$, which is the latent version of $\mathbf{y}$ obtained through the VAE; the resulting signal is subsequently passed through the trainable copy and the second zero-initialized convolutional layer and finally added to the output of the frozen layer, derived from input $\mathbf{y_c}$. The visualization of this process is depicted in Fig. \ref{fig:stable-foley}.

Finally, we use a $\mathrm{depth\_factor}$ parameter to use only a subset of the pre-trained layers of the original DiT: in our experiments we use a $\mathrm{depth\_factor}$ of 20\%, which corresponds to using only 5 layers of the Stable Audio DiT. 

\subsubsection{Semantic Control}

Several recent deep learning investigations have focused on developing versatile audio representations that can be effectively generalized across various downstream tasks \cite{Niizumi2021BYOLFA}. Contrastive learning, in particular, gained great popularity for the training of multimodal models. A notable example of this approach is CLAP \cite{laionclap2023}, which aligns embeddings for both audio and text in a shared latent space. We then condition our sound synthesis model on CLAP audio embeddings during training, enabling it to incorporate text-based conditioning as an additional modality at inference time only.

To further refine the semantic and temporal alignment of the final audio, we also condition the generation process with a direct frame representation of the reference video. We do so using the embeddings provided by the CAVP video encoder, introduced in \cite{NEURIPS2023_98c50f47}. This video encoder aligns the audio and video modalities, providing useful information about the timing and semantics of the related audio. CAVP takes as input video at 4 \textit{fps}, so we use this frame rate to pass only RGB frames to it. 
These semantic controls are fed via cross-attention layers, as global conditionings for Stable Audio. 

Conditionings on duration, represented by the hyperparameters  $\mathrm{seconds\_start}$ and  $\mathrm{seconds\_total}$, indicating the total length of the audio, are fed to the model via prepending it to the input of the model.
Prepend conditioning also includes timestep conditioning, indicating the current diffusion timestep. These conditionings are concatenated along the channels dimension to the input. 

\section{Experimental Setup}
\subsection{Datasets}
\subsubsection{Greatest Hits}

We use the \textit{Greatest Hits} dataset \cite{Owens2015VisuallyIS}, a widely-adopted datasets for V2A tasks.
This dataset includes videos of humans using a drumstick to hit or rub objects or surfaces. The choice of a drumstick as
the striking object is useful, as it minimally occludes each frame, enabling the video model to better comprehend motions in the scene. Each video in the dataset captures the drumstick strokes; the audio is recorded with a shotgun microphone attached to the camera, and then denoised. The dataset provides metadata associated to each video. We use it to define textual prompts according to a predetermined structure: ``A person \textit{\{action\}} \textit{\{frequency\}} on \textit{\{material\}} with a wooden stick", where \textit{\{action\}} -- which can be ``hit" or ``scratch" --, \textit{\{frequency\}} -- which can be either ``multiple times" or ``once" -- and \textit{\{material\}} are derived from the metadata.

The high quality of the samples in this dataset is critical for training V2A models, as very often datasets of video in the wild do not guarantee either good video and audio quality and sufficient audiovisual alignment to allow the models to learn the relationships required to generate audio that is semantically and temporally aligned with the reference video.
Altogether, the dataset consists of 977 videos captured both outdoor and indoor. Indoor scenes contain a variety of hard and soft materials, such as metal, plastic, cloth, while the outdoor scenes contain materials that scatter and deform, such as grass, leaves and water. On average, each video contains 48 actions, divided between hitting and scratching. This ensures that each extracted chunk of video, lasting either 10 seconds, contains a sufficient number of hits.
We divide the dataset into 732 videos for the training set, 49 for
the validation set and 196 for the test set.

\subsubsection{Walking The Maps}

We want to train and test our model on a case study of interest, among the most relevant in the sound design of audiovisual works: the generation of footsteps sounds. Datasets properly adapted for V2A tasks do not exist or are not currently publicly available. A dataset for V2A models must consist of high-resolution video and sound design-quality audio. 
We therefore decided to build a novel dataset with these characteristics in order to test our model in real-world scenarios for Foley synthesis. The audio and video quality of the most modern video games is extremely high, making them the perfect source, in our opinion, to create a V2A dataset. 
Then, we collected clips taken from publicly available YouTube videos of walkthroughs of some famous video games. 

Many gamers upload videos to YouTube in which they roam with their animated character in the maps of different video games. These videos are the perfect target for our dataset; in fact, the sound of footsteps is clearly audible and each footstep is strongly characterized, e.g. steps on grass, concrete or wood sound completely different from each other as well as steps related to a slow walk or a run. High-definition video is associated with such sounds for more modern video games, allowing for an excellent temporal and semantic relationship between audio and video.

\begin{figure}[t]
    \centering
    \includegraphics[width=0.95\linewidth]{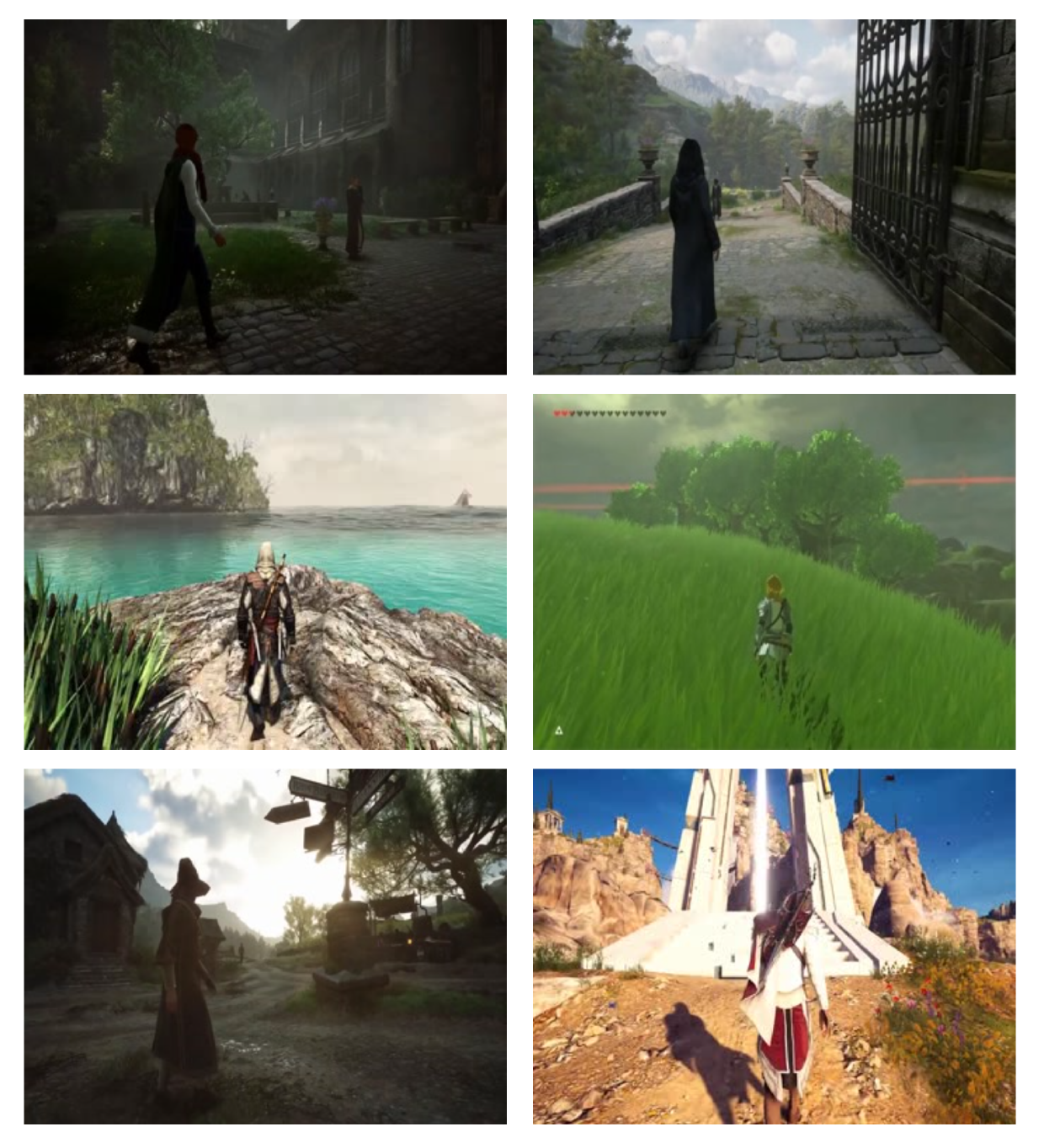}

    \caption{Samples from the \textit{Walking The Maps} dataset highlighting different lighting conditions and grounds.}
    \label{fig:wtm}
\end{figure}

In order to create the current version of our dataset, called \textit{Walking The Maps}, we chose 4 video games as targets: \textit{Hogwarts Legacy}, \textit{Zelda Breath of the Wild}, \textit{Assassin's Creed: Odyssey}, \textit{Assassin's Creed IV Black Flag}.
For each selected video, we extracted only clips in which the sound of the steps is clearly audible and there are no other possible sound sources in the video that can be related to the target sound.

Thus, our dataset is finally composed of 893 video clips of different lengths. The average duration of the videos is 8.82 seconds, where the shortest video has a duration of 2.04 seconds and the longest one is 72.05 seconds long.
Each chunk is saved by reporting the unique ID of the video posted on YouTube, the start second of the chunk related to the full video, and the end instant expressed in seconds, so each video in the dataset will have a name of the type $\mathrm{ID\_start\_end}$.mp4.
While YouTube game walkthroughs often feature prominent footstep sounds, they can also contain background music, dialogue, or other game sounds. To ensure the dataset provides clean target audio primarily containing footsteps, essential for training robust V2A models, we preprocessed the audio from each clip using the AudioSep \cite{liu22_winterspeech} source separation model with the query ``\textit{footstep sounds}".

Since the shortest video in our dataset is 2 seconds long, we fine-tuned our FOL·AI model on 2 seconds chunks of each video in the dataset. The dataset is made publicly available so that it can be used for evaluation of V2A models and to be further extended. 

Some samples from our dataset are provided in Fig. \ref{fig:wtm}

\subsection{Training and Inference Details}

We train the video model and the audio synthesis model separately. 
The video model is trained on a single 48 GB Nvidia RTX A6000. We use a batch size of 64, training the model for 500 epochs with a learning rate maintained constant at $1 \times 10^{-3}$ and Adam as an optimizer with a weight decay set to $1 \times 10^{-3}$.

To train the audio synthesis model, we use the official Stable Audio Open\footnote{\url{https://github.com/Stability-AI/stable-audio-tools}} repository and the related checkpoint\footnote{\url{https://huggingface.co/stabilityai/stable-audio-open-1.0}} to initialize the weights of our model. We use 44.1 kHz stereo audio as ground truth for experiments with \textit{Greatest Hits}. 
the audio synthesis model is trained on a single 48 GB Nvidia RTX A6000 with a batch size of 12 for 20k steps. We use a fixed learning rate of $1 \times 10^{-4}$ and AdamW as an optimizer, with parameters set as it is done in Stable Audio Open. 

In inference, we use the RMS envelopes predicted though the the video model; after $\mu$-law decoding\footnote{\url{https://librosa.org/doc/main/generated/librosa.mu_expand.html}}, we use interpolation to match the sample rate and employ these envelopes as input to the ControlNet of the audio synthesis model. The model then performs 150 sampling steps to generate the final output, using classifier-free guidance \cite{ho2021classifierfree} with a scale set to 2. 

Regarding the experiments on our \textit{Walking The Maps} dataset, we fine-tuned our model starting from the best checkpoints obtained after training on \textit{Greatest Hits} for both the video model and the audio synthesis model. 
The use of 2 second-long chunks implies a change in the temporal dimensions of the data in our model.
Other than that, the parameters used for these experiments are the same as those used for \textit{Greatest Hits}. 


In this case, the video model is fine-tuned with a decreasing learning rate scheduler, with $\gamma=0.5$, for 2500 epochs. While the audio synthesis model is fine-tuned without any change with respect to \textit{Greatest Hits} experiments.

\subsection{Evaluation Metrics}

To perform an objective evaluation of our model, we employ some of the most widely used metrics to attest semantic quality and time alignment in V2A tasks. We use E-L1, acc@1, acc@5, acc@10 for the evaluation of the the video model, while we use FAD-P, FAD-C, FAD-CL, CLAP-score, FAVD and E-L1 for the evaluation of both the audio synthesis model and the complete model, FOL·AI. 

\begin{table*}[!t]
\centering
\caption{Results for FOL·AI and comparison with other SOTA models on \textit{Greatest Hits}. Table shows whether the model generates the output conditioned on an audio or text prompt; HRC stands for Human Readable Control and refers to the use of time-varying interpretable signals that sound designers can use to control the generation process (i.e., envelope or onsets). Our model provides the best results, even in the setting of text conditioned generation.}
\label{tab:overall-results}
\resizebox{\textwidth}{!}{
\begin{tabular}{l|cc|c|c|cccccc}
\toprule
 \textbf{Model} & \textbf{Audio} & \textbf{Text } & \textbf{HRC} & \textbf{FAD-P ↓} & \textbf{FAD-C ↓} & \textbf{FAD-LC ↓} & \textbf{E-L1 ↓} & \textbf{CLAP ↑} & \textbf{FAVD ↓} \\
\midrule
    SpecVQGAN \cite{SpecVQGAN_Iashin_2021} & \ding{55} & \ding{55} & \ding{55}  & 99.07 & 1001 & 0.7102 & 0.0427 & 0.1418 & 6.5136 \\
    
    Diff-Foley \cite{NEURIPS2023_98c50f47} & \ding{55} & \ding{55} & \ding{55} & 85.70 & 654 & 0.469 & 0.0448 & 0.3733 & 4.6186 \\
    CondFoleyGen \cite{du2023conditional} & \ding{51} & \ding{55} & \ding{55} & 74.93 & 650 & 0.4883 & 0.0357 & 0.4879 & 6.4814 \\
    \midrule
\multirow{2}{*}{SyncFusion \cite{Comunit2023SyncfusionMO}}
    & \ding{55} & \ding{51} & \ding{51} & 35.64 & 591 & 0.4365 & 0.0231 & 0.5154 & 4.3020 \\
    & \ding{51} & \ding{55} & \ding{51} & 27.85 & 542 & 0.2793 & 0.0177 & 0.6621 & 3.2825 \\
    \midrule
    \multirow{2}{*}{Video-Foley \cite{Lee2024VideoFoleyTV}} & \ding{55} & \ding{51} & \ding{51} & 67.04 & 644 & 0.4997 & 0.0242 & 0.3680 & 4.9106 \\
    & \ding{51} & \ding{55} & \ding{51} & 28.45 & 435 & 0.1671 & 0.0183 & 0.6779 & 2.2070 \\

\midrule
\multirow{2}{*}{FOL$\cdot$AI (Ours)}
    & \ding{55} & \ding{51} & \ding{51} & 32.80 & 381 & 0.2516 & \textbf{0.0137} & 0.4806 & 3.9413 \\
    & \ding{51} & \ding{55} & \ding{51} & \textbf{16.57} & \textbf{217} & \textbf{0.1048} & \textbf{0.0137} & \textbf{0.6833} & \textbf{2.0264} \\
\bottomrule
\end{tabular}
}
\end{table*}

    \subsubsection{E-L1} Time alignment is evaluated with the \textbf{E-L1} metric introduced in T-Foley \cite{Chung2024TFoleyAC}. E-L1 evaluates the fitting of the generated sounds to the temporal condition of the event:
    \begin{equation}
        E\text{-}L1 = \frac{1}{k} \sum_{i=1}^{k} \|\mathbf{r}_i - \hat{\mathbf{r}}_i\|,
    \end{equation}
    where $\mathbf{r}_i$ is the ground-truth envelope of the i-th frame, and $\hat{\mathbf{r}}_i$ is
    the predicted one.

    \subsubsection{Accuracy metric} We also use \textbf{class-wise accuracy} at $k$ ($k=\{1,5,10\}$). 
    Since we are not trying to make a perfect classification of the various frames in the envelope, as we do not want to penalize near-correct predictions, the most informative accuracy is acc@5. 

    \subsubsection{Fréchet Audio Distance} \textbf{FAD} \cite{KilgourZRS19} is used to evaluate the quality and realism of generated audio compared to reference audio. This metric computes the similarity between the statistical distributions of embeddings of the real and generated audio.

    The choice of audio encoder from which to extract  embeddings significantly impacts the FAD score because different features representations encode specific aspects of audio, so the audio quality measured by this metric with respect to human perception is embeddings dependent \cite{Tailleur2024CorrelationOF}. For this reason we measure FAD using three different audio encoders: PANNs wavegram-logmel \cite{Kong2019PANNsLP} (\textbf{FAD-P}), Microsoft CLAP \cite{Elizalde2023CLAPLA} (\textbf{FAD-C}) and Laion-CLAP \cite{laionclap2023} (\textbf{FAD-LC)}.

    To calculate these metrics, we use the \textit{fadtk}\footnote{\url{https://github.com/DCASE2024-Task7-Sound-Scene-Synthesis/fadtk}} library.

    \subsubsection{CLAP-score metric} \textbf{CLAP-score} is another metric used to assess the overall quality of the generated waveforms, as done in \cite{Evans2024StableAO}. We generate embeddings through CLAP \cite{laionclap2023} of both ground truth and generated audio and compute cosine similarity between them. Since our model is based on the use of CLAP as the audio representation, this metric is useful for attesting how relevant the conditioning audio features are in generating the final output.  

    \subsubsection{Fréchet Audio-Visual Distance} \textbf{FAVD} \cite{10.1007/978-3-031-72986-7_17} is gaining increasing popularity in the evaluation of V2A models, as the purpose of this metric is to measure the temporal and semantic alignment between video and audio modalities. It does so by calculating the Frèchet Distance between video embedding and audio embedding. In our case, we use video and audio encoders used in the reference library, namely I3D \cite{Carreira2017QuoVA} and VGGish \cite{Hershey2016CNNAF}, to calculate the embedding of the ground truth video and the embedding of the generated audio. 


\section{Results}

\subsection{Complete Model}

We compare our complete model, FOL·AI, with the main V2A models publicly available at the time of writing this paper. More precisely, our baseline models are the aforementioned SpecVQGAN \cite{SpecVQGAN_Iashin_2021}, CondFoleyGen \cite{du2023conditional}, Diff-Foley \cite{NEURIPS2023_98c50f47}, Video-Foley \cite{Lee2024VideoFoleyTV} and SyncFusion \cite{Comunit2023SyncfusionMO}. 

For all available models, we use the official code provided on GitHub along with their checkpoints. For comparison with \cite{Lee2024VideoFoleyTV}, we used our video model to generate the RMS envelopes, as at the time of writing this paper the only code and checkpoints available are for the audio synthesis model. 

As shown in Table \ref{tab:overall-results}, FOL·AI achieves improved results across all metrics, including both time alignment and the semantic quality of the generated audio, compared to all baseline models considered. These results highlight the importance of using a strong temporal control, such as the envelope, to guide the generation of audio. Indeed, the second- and third-best models in our tests, Video-Foley and SyncFusion, use envelope and onset tracks for temporal conditioning, providing a stronger guide for temporal alignment than the methods employed by SpecVQGAN, CondFoleyGen, and Diff-Foley.


Furthermore, the results on audio quality metrics demonstrate the critical importance of leveraging a state-of-the-art model like Stable Audio for audio generation. Achieving high audio definition in V2A models is crucial to making these tools actually useful for sound designers.  

\begin{table*}[t!]
\centering
\caption{Results for the video model with different RGB and opt. flow encoders.}
\label{tab:alignment-accuracy}
\resizebox{0.7\textwidth}{!}{
\begin{tabular}{l|cc|ccccc}
\toprule
\textbf{Encoder} & \textbf{RGB} & \textbf{Opt. Flow}& \textbf{E-L1 ↓} & \textbf{Acc@1 ↑} & \textbf{Acc@5 ↑} & \textbf{Acc@10 ↑} \\

\midrule
 BN-Inception & \ding{51}
 & \ding{51}
 & 0.01 & 0.092 & 0.398 & 0.630 \\
TC-CLIP & \ding{51}
 & \ding{55} & 0.0130 & 0.073 & 0.340 & 0.536  \\
TC-CLIP & \ding{51}
 & \ding{51} & \textbf{0.0115} & \textbf{0.104} & \textbf{0.430} & \textbf{0.650} \\

\bottomrule
\end{tabular}
}
\end{table*}

\begin{table*}[t!]
\centering
\caption{Results for the audio synthesis model for audio prompt and ground-truth RMS with and without frames conditioning.}
\label{tab:stablefoley}
\resizebox{0.8\textwidth}{!}{
\begin{tabular}{lccccccc}
\toprule
\textbf{Model} & \textbf{Frames Cond.} & \textbf{FAD-P ↓} & \textbf{FAD-C ↓} & \textbf{FAD-LC ↓} & \textbf{E-L1 ↓} & \textbf{CLAP ↑} & \textbf{FAVD ↓} \\

\midrule
\multirow{2}{*}{GT RMS} & \ding{55} & 9.0722 & \textbf{252} & 0.0635 & 0.0062 & 0.7640 & 1.7103 \\
 & \ding{51} & \textbf{7.67} & 260 & \textbf{0.0630} & \textbf{0.0060} & \textbf{0.7762} & \textbf{1.5832} \\
 
\bottomrule
\end{tabular}
}
\end{table*}

\begin{table*}[h]
\centering
\caption{Results for FOL·AI on \textit{Walking The Maps}.}
\label{tab:video2audio_wtm}
\resizebox{0.56\textwidth}{!}{
\begin{tabular}{lcccccc}
\toprule
\textbf{Model} & \textbf{FAD-C ↑} & \textbf{FAD-LC ↑} & \textbf{E-L1 ↑} & \textbf{CLAP ↑} & \textbf{FAVD ↓} \\

\midrule
FOL·AI & 167 & 0.2556 & 0.0460 & 0.6956 & 3.0682 \\
 
\bottomrule
\end{tabular}
}
\end{table*}

The use of ControlNet enables us to leverage the prior knowledge of Stable Audio for generating ambient sounds, allowing the production of 44.1 kHz stereo audio effects, matching professional audio production standards. In fact, compared with the baseline methods reported, ours is the only one that generates audio at a standard rate for professional audio. Additionally, the integration of ControlNet with a high $\mathrm{depth\_factor}$ of 20\% ensures that our model is both lightweight and fast. Indeed, the results we achieve are based on training the model on a small dataset of approximately 6 hours, such as \textit{Greatest Hits}, with a limited number of training steps, thus avoiding the need for substantial time and computational resources.

Table \ref{tab:video2audio_wtm} shows the results on objective metrics obtained by evaluating our model on the \textit{Walking The Maps} test set. Both the video model and the audio model, succeed to obtain good results for the synchronization between audio and video as well as for the quality of the produced waveforms, managing to generate realistic footsteps sounds. The sounds produced are diversified from each other according to both the character's walking style and the ground type.

\subsection{Ablation studies} 
\subsubsection{Different encoders in the video model} Our experiments demonstrate the fundamental importance of the optical flow information. When we provide the model with only the features derived from video frames, the results are significantly worse compared to using the concatenation of RGB and optical flow features.  

We also experiment with the feature extractor used in RegNet, that is BN-Inception \cite{10.5555/3045118.3045167}. This network is trained on images, so the features extracted for one frame are not correlated with those of other frames. However, it is trained on an extremely large dataset (ImageNet \cite{5206848}), allowing it to generalize well and produce strong results even when encoding video frames, despite not being designed for this purpose. Our experiments show that TC-CLIP achieves better results than BN-Inception.
This outcome is significant, as TC-CLIP is trained on a smaller dataset (Kinetics-400 \cite{Kay2017TheKH}) specifically designed for motion recognition. In our opinion, this demonstrates the direction V2A research should pursue: employing models specifically designed to capture temporal context.  

Results for the video model and related ablation studies are reported in Table \ref{tab:alignment-accuracy}.

\subsubsection{Use of video features in the audio synthesis model} We evaluate the audio synthesis model by generating audio through ground truth RMS envelopes extracted directly from waveforms. As ablation study, we train the model without the direct conditioning of frames provided by CAVP. The results reported in Table \ref{tab:stablefoley} show that such additional information leads to a perceptible improvement in the semantics of the produced audio, as can be noticed from the metrics for semantic alignment. While there is no perceivable enhancement in temporal alignment; this result shows that temporal conditioning is contributed entirely by the ControlNet input, demonstrating the strength of such conditioning, which is the desired result. 

\section{Conclusion and Discussion}

In this paper, we present a novel model for generating an audio track semantically and temporally aligned to a silent input video, called FOL·AI. The model is divided into two distinct parts, trained separately and joined only at inference time: the video model, which maps a representative envelope of the audio to be generated directly from the reference video, and the audio synthesis model that, through the use of the predicted envelope and other semantic controls, generates the final output of the model. 
Our sound synthesis model leverages Stable Audio Open and, to the best of our knowledge, this is the first time that such a state-of-the-art model for audio generation is used in the context of Video-to-Audio. The main contributions of FOL·AI also include the novel application of ControlNet to Stable Audio's DiT architecture for precise temporal control using RMS envelopes, the introduction of the \textit{Walking The Maps} dataset focused on the challenging Foley task of footstep synthesis, and achieving state-of-the-art performance in V2A tasks while generating high-fidelity 44.1 kHz stereo audio.


Although the audio synthesis model tracks with high accuracy the envelope used as a control, mapping that envelope from the video represents the bottleneck of our model, still leaving room for potential future research works. 

In addition, datasets suitable for V2A problems, containing high-quality video and audio, are still too few or difficult to obtain. In our case, we believe that the introduction of suitable datasets, such as \textit{Walking The Maps}, can be of great help for researchers in the field of Foley synthesis. 

Also, the semantics of the generated waveforms can be further optimized through the use of different and more relevant semantic conditionings. 

\section{Acknowledgements}

This work was supported by the European Union under the Italian National Recovery and Resilience Plan (NRRP) of NextGenerationEU, partnership on “National Centre for HPC, Big Data and Quantum Computing” (CN00000013 - Spoke 6: Multiscale Modelling \& Engineering Applications). 

This work was supported by “Progetti di Ricerca Medi” of Sapienza University of Rome for the project “SAID: Solving Audio Inverse problems with Diffusion models”, under grant number RM123188F75F8072.

E.P. and L.C. were partially supported by PRIN 2022 project no. 2022AL45R2 (EYE-FI.AI, CUP H53D2300350-0001).
 
M. C. was funded by UKRI and EPSRC as part of the “UKRI CDT in Artificial Intelligence and Music”, under grant EP/S022694/1.

\bibliographystyle{IEEEbib}
\bibliography{ref}

\end{document}